\documentclass[twocolumn,showpacs,preprintnumbers,amsmath,amssymb,prb,
superscriptaddress]{revtex4}

\usepackage{graphicx}   % Include figure files
\usepackage{dcolumn}    % Align table columns on decimal point
\usepackage{bm}         % bold math
\begin{document}

\title{Generation of EPR pairs and interconversion
of static and flying electron spin qubits}

\author{G. Giavaras}
\altaffiliation[Present address: ]{Department of Physics and
Astronomy, University of Leicester, Leicester, England}
\affiliation{Department of Physics, Lancaster University,
Lancaster LA14YB, England} \affiliation{QinetiQ, St.Andrews Road,
Malvern WR143PS, England}
\author{J. H. Jefferson}
\affiliation{QinetiQ, St.Andrews Road, Malvern WR143PS, England}
\author{M. Fearn}
\affiliation{QinetiQ, St.Andrews Road, Malvern WR143PS, England}
\author{C. J. Lambert}
\affiliation{Department of Physics, Lancaster University,
Lancaster LA14YB, England}

\date{\today}

\begin{abstract}
We propose a method of generating fully entangled electron spin
pairs using an open static quantum dot and a moving quantum dot,
realised by the propagation of a surface acoustic wave (SAW) along
a quasi-one-dimensional channel in a semiconductor
heterostructure. In particular, we consider a static dot (SD)
loaded with two interacting electrons in a singlet state and
demonstrate a mechanism which enables the moving SAW-dot to
capture and carry along one of the electrons, hence yielding a
fully entangled static-flying pair. We also show how with the same
mechanism we can load the SD with one or two electrons which are
initially carried by a SAW-induced dot. The feasibility of
realizing these ideas with existing semiconductor technology is
demonstrated and  extended to yield flying or static pairs that
are fully entangled and arbitrary interconversion of static and
flying electron spin qubits.
\end{abstract}

\maketitle

Einstein-Podolsky-Rosen (EPR) particle pairs are spatially
separated and fully entangled particles such as photons or
electrons~\cite{einstein}. A spin entangler system, which is a
system capable of generating EPR spin pairs, is of fundamental
importance not only in the field of quantum computation but also
in quantum mechanics for testing nonlocal
correlations~\cite{einstein}. In addition the ability to
interconvert static and flying qubits is a basic requirement for
quantum computation and communication in general~\cite{divinc}. In
the solid state both the creation of an entangler for massive
particles and interconversion of qubits are hard tasks since the
process of generation and detection must take place in a
controlled way and in a much shorter time than typical decoherence
times. Electron spin qubits in semiconductors are particularly
promising since their decoherence times can be quite long, e.g.
$\sim$100 ns in GaAs~\cite{kikkawa}, and a number of entangler
schemes have been studied, at least theoretically. These for
example include EPR pairs via Coulomb scattering~\cite{saraga1}, a
triple quantum dot structure~\cite{ping,saraga2}, a three-port
dot~\cite{oliver}, a superconductor attached to a double
dot~\cite{recher1} and a turnstile double dot
device~\cite{burkard,hu,blaauboer,legel}.

\begin{figure}
\begin{center}
\includegraphics[width=7.3cm,height=5cm]{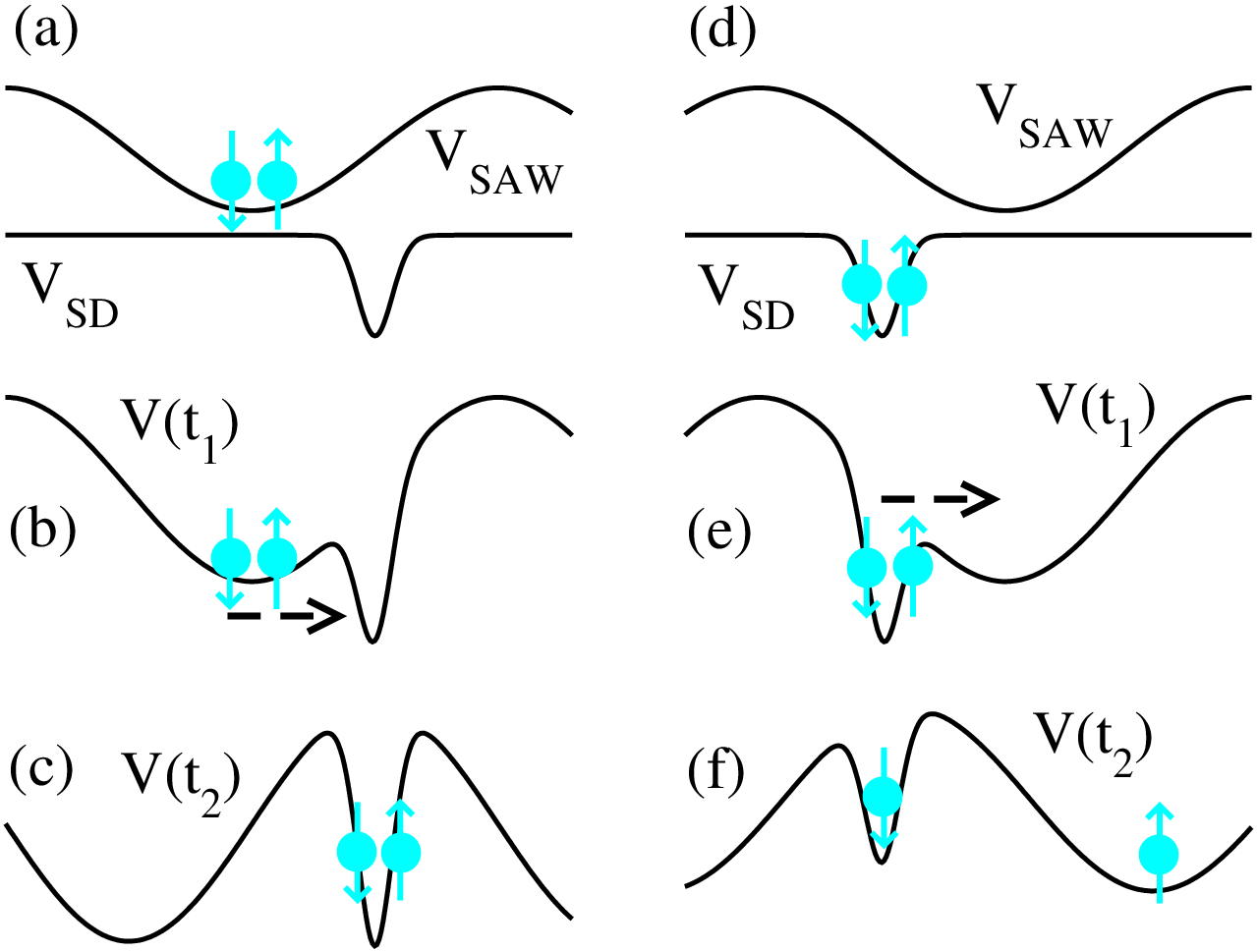}
\caption{(color on line). Schematic illustration of the proposed
system to generate EPR electron spin pairs. (a) An empty static
dot (SD) with a confining potential, $V_{SD}$, is formed in a
depleted GaAs/AlGaAs one-dimensional channel. A surface acoustic
wave (SAW) induces a sine-wave potential, $V_{SAW}$, which
propagates along the channel (from left to right) forming moving
quantum dots. A SAW-dot is adjusted to contain two electrons which
in general may not be in a singlet state. (b) The total potential
$V=V_{SAW}+V_{SD}$ enables tunneling whose degree can be
efficiently controlled using a time dependent gate voltage which
tunes the depth of the SD. (c) Within a regime of parameters the
two electrons can tunnel from the SAW-dot into the SD and can
remain bound as the SAW propagates (alternatively the electrons
can tunnel into the SD sequentially). Then the electrons are
allowed to relax to their singlet ground state. (d) The SD
contains two electrons in a singlet ground state while the SAW-dot
is empty. (e) Due to the time dependent gate voltage controlled
tunneling can take place from the SD into the SAW-dot. (f) One
electron can tunnel from the SD into the SAW-dot and it can be
carried along by the SAW leaving the second electron trapped in
the SD, hence generating a static-flying EPR pair. The SD-electron
can then be ejected into a SAW-dot that follows generating a
flying EPR pair, or the SAW-electron can be captured by a nearby
SD generating a static EPR pair.}
\end{center}
\end{figure}

As has been demonstrated experimentally the propagation of a
surface acoustic wave (SAW) along a depleted one-dimensional
channel in a GaAs/AlGaAs two-dimensional electron gas (2DEG)
results in the formation of moving quantum
dots~\cite{shilton,talyanskii}. The heterostructure confines the
electrons in the plane of the quantum well, whereas lateral
confinement in one direction is due to the channel and in the
other direction to the induced time dependent electrostatic
potential which accompanies the SAW mechanical propagation. The
moving SAW-induced dots can capture electrons from the 2DEG and
transport them through the channel. Under a regime of parameters
of SAW power and gate voltage the number of transported electrons
by the SAW can be controlled down to one yielding a quantized
acoustoelectric current of nA with an accuracy of five parts in
10$^{4}$.~\cite{shilton,talyanskii,cunningham} In general, this
current can be expressed as $nef$, where $n=1,2...$, is the
average number of electrons in the SAW-dots, $f\sim2.7$ GHz is the
SAW frequency and $e$ is the absolute electronic charge. Spin
qubits in SAW-dots have been investigated as a possible
realisation of a quantum computer~\cite{barnes1,furuta,gumbs} and
recently it was demonstrated theoretically that maximal
entanglement can arise from the interaction between a SAW spin
qubit and a spin qubit trapped in a static dot (SD), embedded in
the depleted channel along which the SAW
propagates~\cite{giavaras}. Experimentally the interaction between
SAW-dots and a SD has been examined mainly in terms of the
acoustoelectric current~\cite{ebbecke1,ebbecke2,fletcher} though
the main goal is the fabrication of a SAW-based device for quantum
information processing~\cite{kataoka}.

In this work we demonstrate the feasibility of generating fully
entangled pairs of electron-spin qubits to very high fidelity.
This differs fundamentally from our earlier work~\cite{giavaras}
for which the degree of entanglement was highly dependent on
initial conditions, with associated uncertainty in a realistic
device. In addition, we demonstrate an efficient mechanism for
interconverting the resulting spin qubits from static to flying
and vice-versa, an issue which has not been addressed in solid
state systems. The basic idea underlying the present proposal is
to use a SAW in conjunction with a synchronised variation in gate
potential to either eject or capture electrons in a SD in a
controlled way. This may be used to generate EPR pairs by first
loading the SD with two electrons which are allowed to relax to
their singlet ground state. As shown schematically in Fig. 1, one
electron is then ejected into a SAW-dot resulting in a delocalized
static-flying pair of fully entangled spin-qubits. By the same
method, the flying qubit may be subsequently captured by another
SD or the static qubit may be ejected into another SAW-dot and in
this way arbitrary interconversion of static and flying spin
qubits may be achieved. The SAW-based loading mechanism of the SD,
which is also efficient for two electrons, may be employed in
static-flying spin qubit interactions~\cite{giavaras,john,costa}

The two-electron problem is studied within the effective mass
approximation by considering the parabolic band Hamiltonian
\begin{equation}\label{Eq.1}
H=\sum_{i=1,2}\left[
-\frac{\hbar^{2}}{2m^{*}}\frac{\partial^{2}}{\partial
x^{2}_{i}}+V(x_{i},t)\right]+V_{c}(x_{1},x_{2}).
\end{equation}
where $m^{*}=0.067m_{o}$ is the effective mass of the electrons in
GaAs. The total potential is given by the sum of the SAW and the
SD time-dependent potentials i.e.,
$V(x,t)=V_{SAW}(x,t)+V_{SD}(x,t)$. The SAW potential can be
modelled efficiently by the form~\cite{giavaras}
$V_{SAW}(x,t)=V_{o}\cos[2\pi(x/\lambda-ft)]$, with the SAW
frequency $f$=2.7 GHz and the SAW wavelength $\lambda$=1 $\mu$m.
These are typical values for SAW-based single electron transport.
Finally, $V_{o}$ is the SAW potential amplitude, which can be
controlled experimentally by the applied power to the transducer
which is used to generate the SAW~\cite{shilton,talyanskii}. The
confining potential of the SD is modelled by the form
$V_{SD}(x,t)=-V_{w}(t)\exp(-x^{2}/2l^{2}_{w})$, where $l_{w}$
determines the width of the dot and $V_{w}(t)$ is the
corresponding depth of the dot which is controlled by a time
dependent gate voltage. Below we describe in detail the time
dependence of $V_w(t)$. We note that the viability of the
mechanism described below is independent of the particular form of
$V_{SD}(x,t)$. In addition, the Gaussian dot that we have chosen
can approximate a parabolic dot well, which is consistent with
experiments on gated quantum dots~\cite{singl}. The Coulomb term
is $V_{c}(x_{1},x_{2})=e^{2}/4\pi \epsilon_{r}\epsilon _{o} r$
with $r=\sqrt{(x_{1}-x_{2})^{2}+\gamma_{c}^{2}}$ and
$\gamma_{c}$=20 nm, ensuring that excited-state modes in $y$ and
$z$ directions have negligible occupation. For GaAs we have taken
$\epsilon_{r}=13$.

To determine the two-electron state of the SD when $V_{SAW}$=0,
and with a time independent SD potential we diagonalise
numerically Hamiltonian (1). The ground state $\Phi(x_{1},x_{2})$
is a singlet, as has been verified experimentally for parabolic
quantum dots~\cite{singl}. This fully spin entangled state has the
orbital distribution $\rho=\int|\Phi(x,x')|^{2}dx'$ shown in the
inset of Fig.~\ref{dot}(a) for the parameters $V_{w}$=13 meV,
$l_{w}=25$ nm. The two electrons are close to the weak correlation
regime~\cite{bryant} (the kinetic energy dominates over the
Coulomb energy) having an orbital distribution which peaks in the
centre of the SD similarly to the non-interacting limit and, more
importantly, the ground state singlet is well separated from
excited states.

The time evolution of $\Phi(x_{1},x_{2})$ when $V_{SAW}$ is
switched on, while keeping $V_{SD}$ fixed in time, is determined
by integrating numerically the time-dependent Schr\"odinger
equation. For $V_{o}$=5 meV and, for example, during one SAW
period, $T=1/f=0.37$ ns, the state remains well-localized in the
SD without tunneling and the electron distribution within the SD
is hardly changed. In particular, the state evolves via
non-adiabatic Landau-Zener transitions (LZTs) which enable the
state to retain its initial character by changing in time the
instantaneous eigenstate index ($n\rightarrow n\pm1$) at each
anticrossing point~\cite{maksym,zener}. To quantify this behaviour
we have determined the instantaneous eigenspectrum
$E_{n}=E_{n}(t)$, $n=0,1$... by solving the eigenvalue problem
$H(t)\Phi_{n}(t)=E_{n}(t)\Phi_{n}(t)$ (for singlet states) at each
instant in time treating $t$ as parameter. The LZT probability at
an anticrossing point with energy gap $2\delta$ is given by an
approximate expression~\cite{maksym,zener} $P_{LZ}\sim
1-2\pi\delta^{2}/\hbar\Lambda$, where the matrix element
$\Lambda=<\Phi_{e}|\partial H/\partial t|\Phi_{l}>$ involves the
states $\Phi_{e}$, $\Phi_{l}$ which enter and leave the
anticrossing point respectively and for the problem under study
typical values are $P_{LZ}>$0.99. The number of anticrossing
points for a fixed $V_{o}$ increases with decreasing $V_{w}$ and
$P_{LZ}$ decrease since the tunnel barrier between SD and SAW-dot
weakens~\cite{giavaras}. In the limit of a very small $V_{w}$ the
initial electron state cannot follow the SAW propagation even for
one cycle, since due to the strong tunneling, it spreads into the
continuum immediately when the SAW interacts with the SD. In the
opposite limit of a very large $V_{w}$ the two-electron state
$\Phi(x_{1},x_{2})$ evolves adiabatically and specifically it
corresponds to the ground state at all times and therefore the
evolution involves no anticrossing points.

\begin{figure}
\begin{center}
\includegraphics[width=7.5cm]{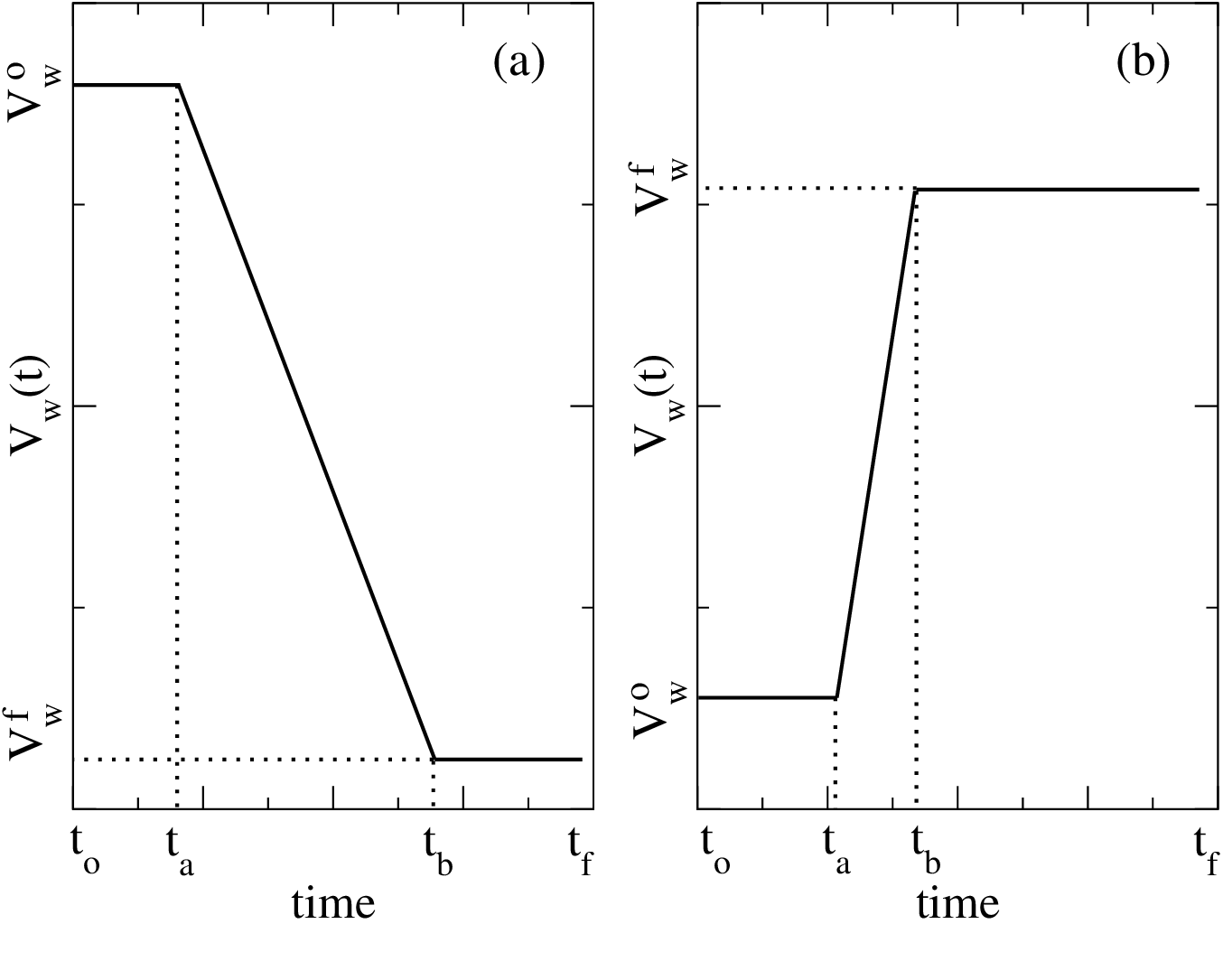}
\includegraphics[width=7.5cm]{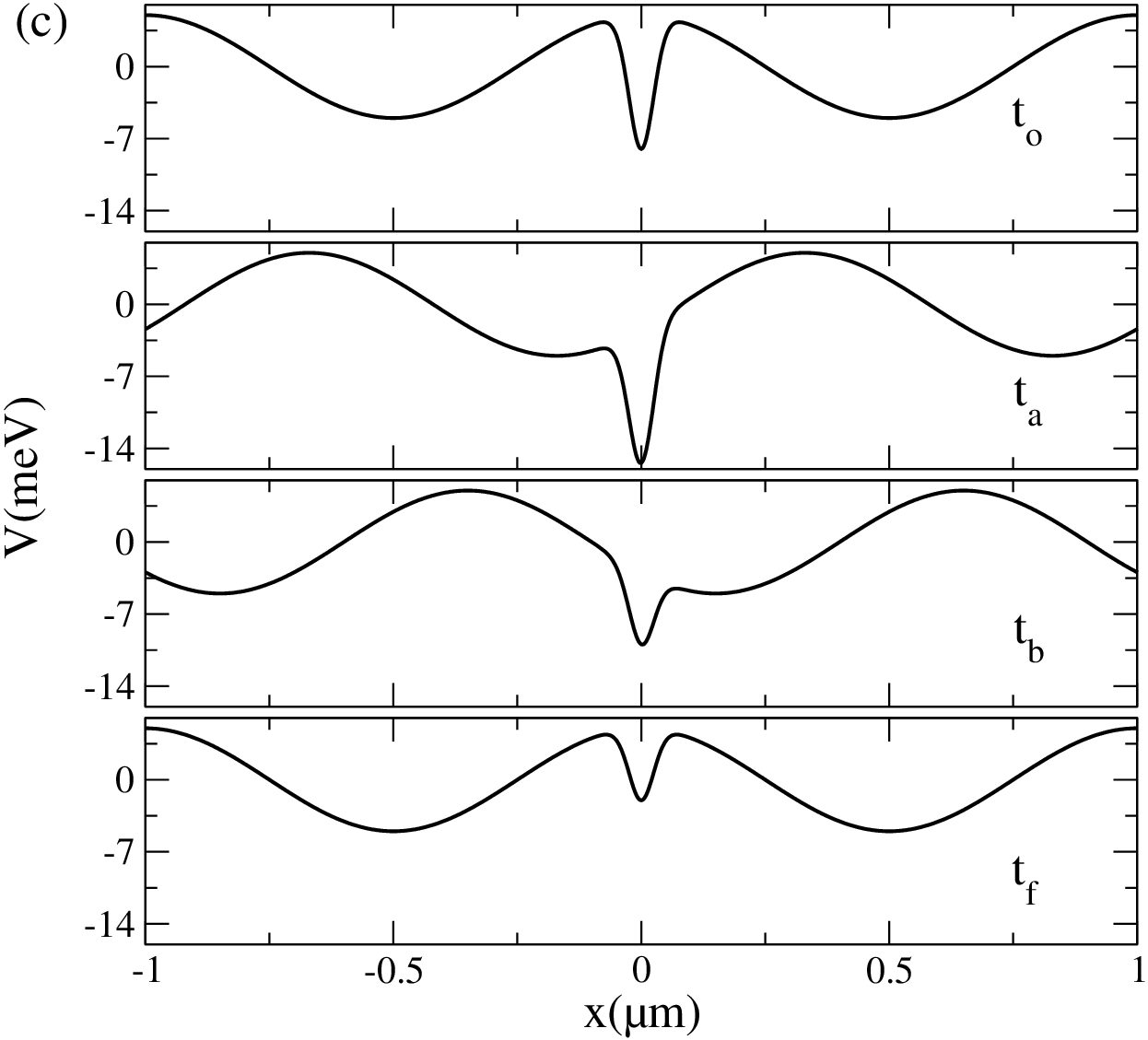}
\caption{(a) Variation of the SD depth, which is due to the time
dependent gate voltage, when electrons are transferred from the SD
to the SAW-dot. (b) The same as in (a) but when electrons are
transferred from the SAW-dot to the SD. In both cases $V^{o}_{w}$
is the initial depth of the SD, $V^{f}_{w}$ is the final depth of
the SD and the tuning takes place from $t_a$ to $t_b$. These
quantities are not in general the same for both processes. (c)
Total potential for four different times when electrons are
transferred from the SD to the SAW-dot. Note that in this case the
electrons are bound at $t_{o}=0$ in the SD ($x=0$) whose depth
decreases in time from $t_a$ to $t_b$. On the other hand, when
electrons are transferred from the SAW-dot to the SD they are
bound at $t_{o}=0$ in the SAW-dot ($x=-0.5$ $\mu$m) and the SD
potential increases in time from $t_a$ to $t_b$. The SAW
propagates always from left to right.}\label{pulse}
\end{center}
\end{figure}

The spin entangler mechanism that we propose using Hamiltonian
(1), which conserves the symmetry of the state in time, is
feasible as long as the resulting total time-dependent potential
changes in such a manner so as to allow one electron to tunnel
from the SD into the SAW-dot during a period of time for which the
LZTs occur with negligible probability. On the other hand, after
that period of time the potential must enable LZTs with high
probability in order that the electron remaining in the SD remains
well-localised as the SAW propagates. We have verified that such
behaviour can be achieved by investigating the problem numerically
for a typical range of parameters. A simple technique to tune the
total potential in this way is to decrease the depth of the SD
with a linear gate voltage ramp. In this work we have used
\begin{equation}
V_{w}(t)=\left\{
\begin{array}{c}
    V^{o}_{w} ,              \qquad       \qquad      \quad               t
\leq t_{a}
\\
\\ V^{o}_{w} - \Delta_{w} \frac{(t-t_{a})}{t_b-t_a}, \qquad   t_{a} \leq t\leq
t_{b}
\\
\\  V^{o}_{w}-\Delta_w ,      \qquad               t  \geq t_{b}
\end{array}
\right.
\end{equation}
where $\Delta_{w}=V^o_{w}-V^f_{w}$. In this expression $V^o_{w}$
is the initial depth of the SD, $V^f_{w}$ is the final depth of
the SD and the tuning takes place from $t_a$ to $t_b$. In
Fig.~\ref{pulse}(a) we show how the variation of the SD depth,
$V_w(t)$, takes place in time when electrons are transferred from
the SD to the SAW-dot. For comparison we also show in
Fig.~\ref{pulse}(b) $V_w(t)$ for the inverse process i.e., when
electrons are transferred from the SAW-dot to the SD a process
that we demonstrate below. The important point is that $V_w(t)$
decreases when electrons are transferred from the SD to the
SAW-dot ($\Delta_{w}>0$), whereas $V_w(t)$ increases when
electrons are transferred form the SAW-dot to the SD
($\Delta_{w}<0$). Figure~\ref{pulse}(c) illustrates the variation
of the total time dependent potential for four different times
when electron transfer from the SD to the SAW-dot takes place. In
this case the electrons are bound at $t_{o}$ in the SD ($x=0$)
whose depth decreases in time from $t_a$ to $t_b$. When electron
transfer from the SAW-dot to the SD takes place the electrons are
bound at $t_{o}=0$ in the SAW-dot ($x=-0.5$ $\mu$m) and the SD
depth increases in time from $t_a$ to $t_b$. Note that in both
cases the SAW propagates from left to right.

\begin{figure}
\begin{center}
\includegraphics[width=8.5cm,height=5.5cm]{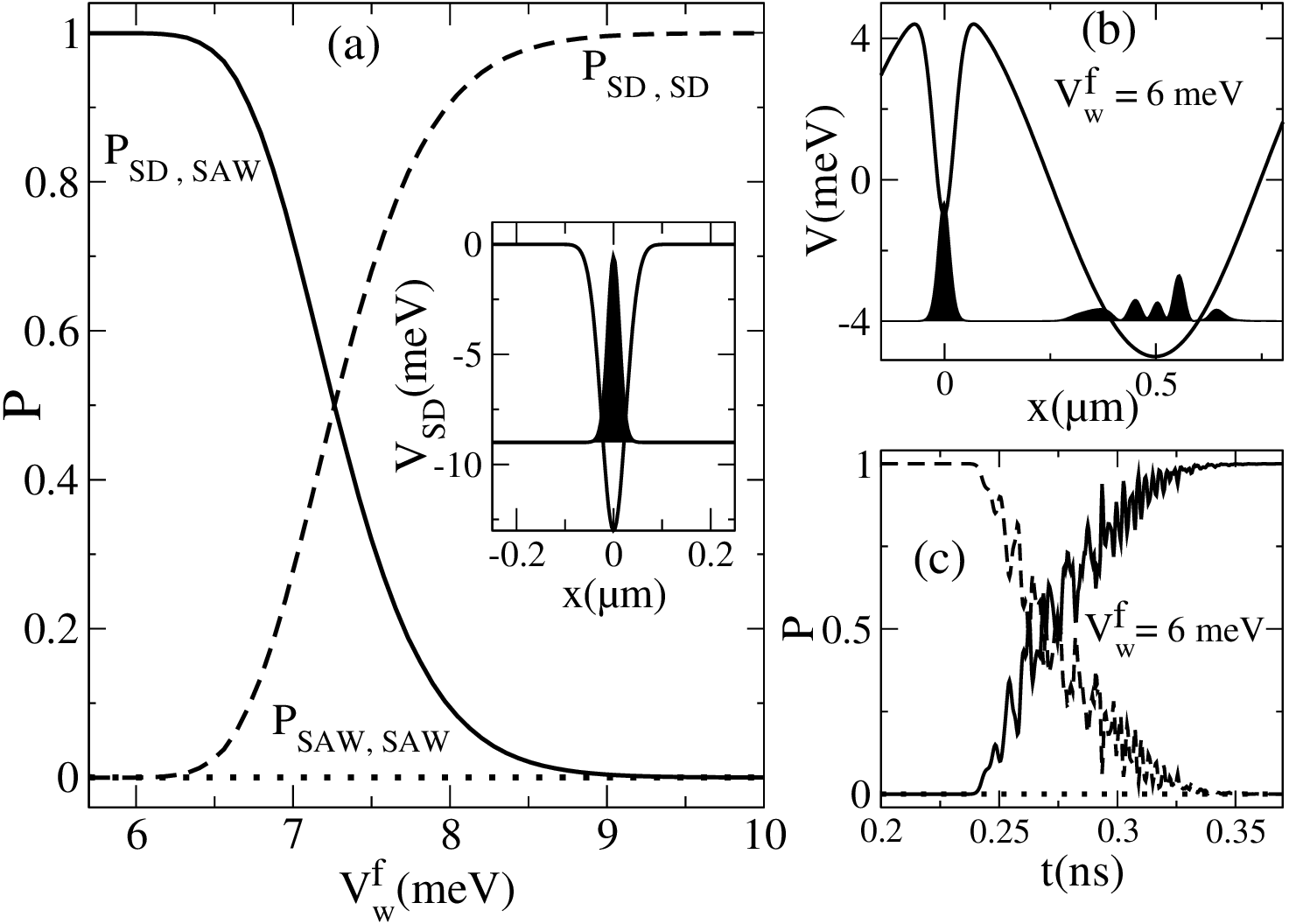}
\caption{(a) Occupation probabilities at the final time
$t_{f}=T=0.37$ ns as a function of the final depth of the SD, when
the initial depth of the SD is $V^o_{w}$=13 meV,
$t_{a}=0.45T=0.17$ ns, $t_{b}=0.65T=0.24$ and at the initial time
$t_{o}=0$ the two electrons are trapped in the SD. $P_{SD,SD}$ is
the probability of finding the two electrons in the SD,
$P_{SAW,SAW}$ is the probability of finding the two electrons in
the SAW-dot and $P_{SD,SAW}$ is the probability of finding one
electron in the SD and the other in the SAW-dot. The inset shows
the initial ($t_{o}=0$) two-electron distribution (in arbitrary
units) in the SD when $V_{SAW}=0$. (b) Total potential and
two-electron distribution at final time $t_{f}=T=037$ ns, when
$V^{f}_{w}$=6 meV and at $t_{o}=0$ the two electrons are trapped
in the SD. (c) The same as in (a) but as a function of time when
$V^{f}_{w}$=6 meV.}\label{dot}
\end{center}
\end{figure}

First we describe the EPR pair generation where at the initial
time $t_{o}=0$ the two electrons are bound in the SD confining
potential. In Fig.~\ref{dot}(a) we show the relevant occupation
probabilities at the final time $t_{f}=T=0.37$ ns for different
final SD depths, $V^{f}_{w}$, but with an initial depth
$V^{o}_{w}$=13 meV for all cases. In this example
$t_{a}=0.45T=0.17$ ns and $t_{b}=0.65T=0.24$ ns. A key observation
for the efficacy of the entangler is that the SD potential,
$V_{SD}$, must change as the potential minimum of the SAW-dot
sweeps past it, otherwise the fraction of the initial state which
tunnels off the SD cannot be trapped and carried along by the
SAW-dot. This requires that $T_p=t_b-t_a$ is of order $T$, which
is within the limit of available electronics. Even though for all
$V^{f}_{w}$ in Fig.~\ref{dot} the SD can bind two electrons when
$V_{SAW}$=0, the physical effect of the time dependent gate
voltage when combined with the SAW potential is to decrease in
time the binding energy of the SD forcing one electron to tunnel
partly or even totally into the SAW-dot. This is why $P_{SD,SD}$
decreases with decreasing $V^{f}_{w}$ whilst $P_{SD,SAW}$
increases. In this study the range that we vary $V^{f}_{w}$
ensures that a single electron can successfully accomplish LZTs,
following the SAW propagation without tunneling, therefore
$P_{SAW,SAW}\sim0$. On the other hand, when double occupation in
the SD is appreciable, uncontrollable tunneling off the SD can
still occur after $t_{f}=T=0.37$ ns. This can be minimized by
restoring the SD potential to its original depth with the gate,
though this regime with appreciable $P_{SD,SD}$ is not appropriate
for the entangler. We can achieve a nearly ideal operation mode
for $V^{f}_{w}\lesssim$6 meV which, as shown in Fig.~\ref{dot}(a),
at the final time $t_{f}=T=0.37$ ns results in one trapped
electron in the SD and a second electron trapped in the moving
SAW-dot with very high probability $P_{SD,SAW}>$0.99. However, we
point out that when the SD depth is made too shallow
($V^{f}_{w}\lesssim$3 meV) both electrons tunnel into the SAW-dot
and $P_{SAW,SAW}=1$. For the case $V^{f}_{w}=$6 meV
Fig.~\ref{dot}(b) shows the final electron distribution when the
Coulomb interaction is negligible and Fig.~\ref{dot}(c) shows the
occupation probabilities as a function of time~\cite{note0}. The
SD-state is the corresponding lowest eigenstate, whereas the
SAW-state consists of a superposition of the lowest SAW-dot
eigenstates whose number can be controlled with choice of $T_{p}$,
SD and SAW parameters. At the final time the SD-state develops via
non-adiabatic LZTs, whereas the SAW-state develops adiabatically,
to a good approximation, with no leakage due to the high
confinement provided by $V_{o}$~\cite{giavaras}.

The electron which remains trapped in the SD can be made to follow
the SAW-electron by further decreasing the depth of the SD and
releasing the trapped electron in a SAW-dot that follows. In this
way we may create flying entangled nonlocal electron pairs
travelling at the sound velocity $\lambda f$=2700 ms$^{-1}$. Since
the wave packets are trapped in the SAW-dots, any physical
spreading to neighboring SAW-dots is negligible. In fact the
overlap between states of neighboring SAW-dots can be controlled
and even completely minimised by choosing a very strong SAW
potential amplitude. If necessary for measurement the two
electrons can be driven to different channels using a Y-branch
design~\cite{valery3} or isolated SAW-pulses~\cite{barnes1}, to
avoid interference effects. Detection and measurement of the
flying SAW-dot spins can be performed with the techniques of
Refs.~16 and ~17. Another option (described below) is to trap the
SAW-electron in a nearby SD using a synchronised time-dependent
gate voltage yielding a nonlocal spin singlet with one electron
each in two spatially delocalized SDs. Measurement of the SD-spins
can then be made using the spin to charge conversion
method~\cite{hanson} and finally the spins can be removed via
tunneling or by a SAW to allow repetition of the cycle. In either
case the empty SD may be loaded with two electrons from leads or
by using a SAW, as we show below, and the whole process repeated.

\begin{figure}
\begin{center}
\includegraphics[width=7cm,height=5cm]{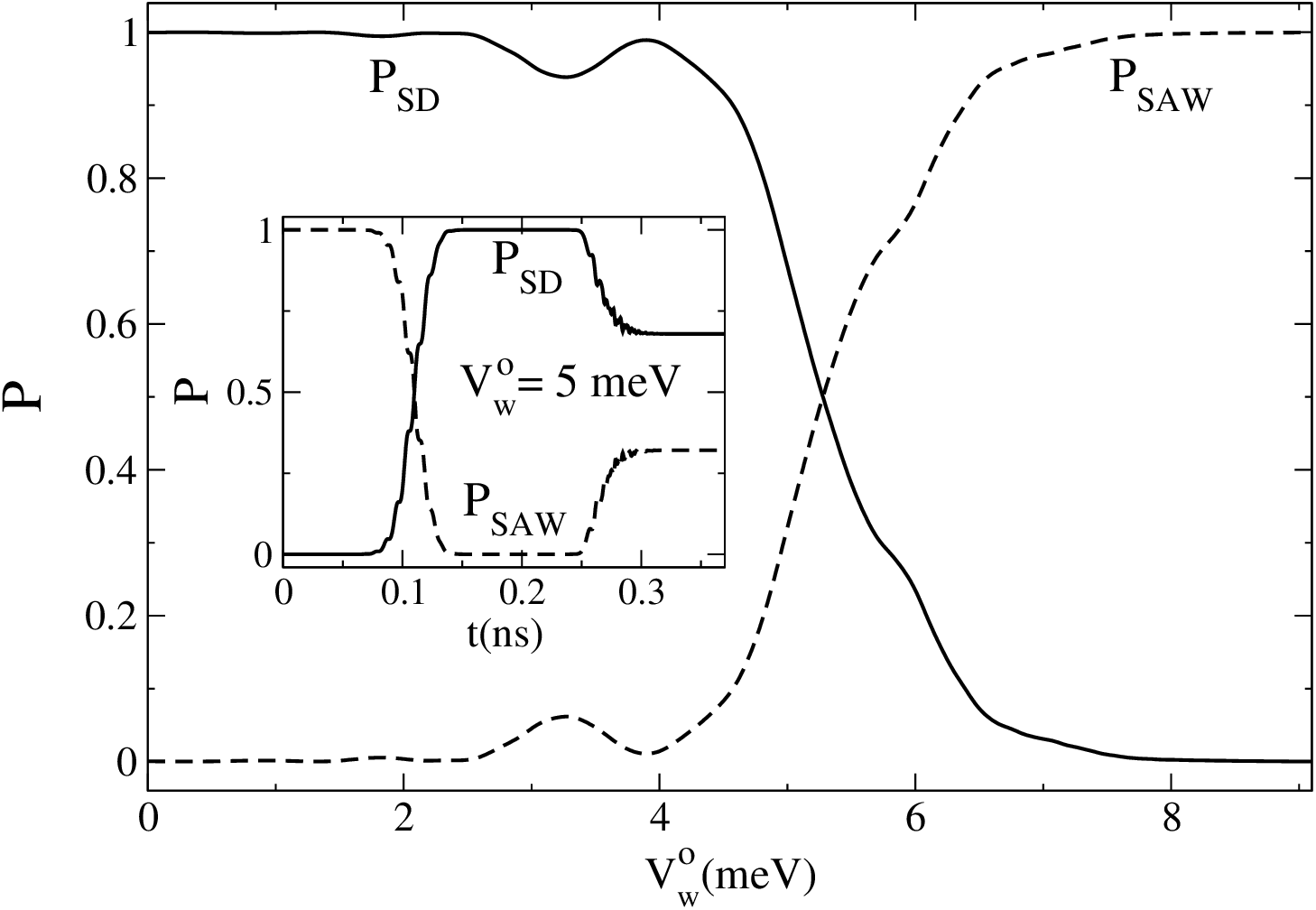}
\caption{Occupation probabilities at final time $t_{f}=T=0.37$ ns
as a function of the initial depth of the SD, when the final depth
of the SD is $V^{f}_{w}$=11 meV, $t_{a}=0.39T=0.14$ ns,
$t_{b}=0.58T=0.21$ ns and at the initial time $t_{o}=0$ one
electron is carried by the SAW-dot. $P_{SD}$ is the probability of
loading the SD with a single electron and $P_{SAW}$ is the
probability of keeping the electron in the SAW-dot. The inset
shows a typical example of the time evolution when $V^{o}_{w}=5$
meV.}\label{single}
\end{center}
\end{figure}

\begin{figure}
\begin{center}
\includegraphics[width=8.5cm,height=5.5cm]{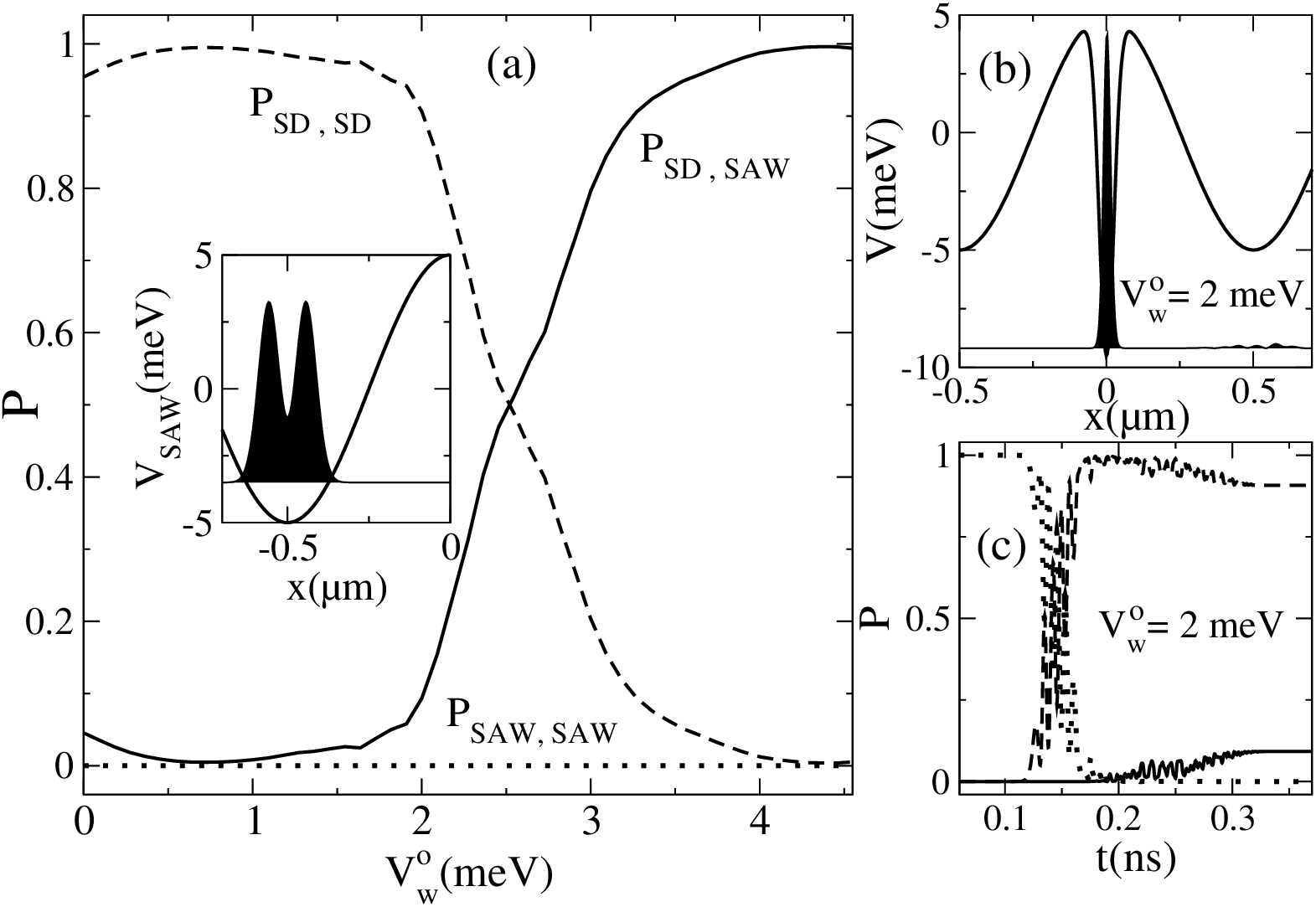}
\caption{(a) Occupation probabilities at the final time
$t_{f}=T=0.37$ ns as a function of the initial depth of the SD
when the final depth of the SD is $V^{f}_{w}$=14.5 meV,
$t_{a}=0.42T=0.16$ ns, $t_{b}=0.7T=0.26$ ns and at the initial
time $t_{o}=0$ the two electrons are carried by the SAW-dot.
$P_{SD,SD}$ is the probability of loading with two electrons the
SD, $P_{SAW,SAW}$ is the probability of keeping the two electrons
in the SAW-dot and $P_{SD,SAW}$ is the probability of loading with
a single electron the SD, while the second electron remains in the
SAW-dot. The inset shows the initial ($t_{o}=0$) two-electron
distribution (in arbitrary units) in the SAW-dot. (b) Total
potential and two-electron distribution at final time
$t_{f}=T=037$ ns, when $V^{o}_{w}$=2 meV and at $t_{o}=0$ the two
electrons are carried by the SAW-dot. (c) The same as in (a) but
as a function of time when $V^{o}_{w}$=2 meV.}\label{saw}
\end{center}
\end{figure}

We now present results on the loading process of a SD with one or
two electrons using a moving SAW-dot. The former process enables
spin conversion from flying to static, while the latter can be the
starting point for EPR pair generation described above. For this
purpose a SAW-dot carrying at $t_{o}=0$ one or two
electrons\cite{note1} is driven towards the empty SD whose
confining potential is increased synchronously with a time
dependent gate voltage according to Eq.(2). Specifically, in
Fig.~\ref{single} we show the relevant occupation probabilities at
the final time $t_{f}=T=0.37$ ns (similar behaviour is displayed
for $t>t_{f}$) as a function of $V^{o}_{w}$ for a single electron
which at $t_{o}=0$ is in the ground state of the SAW-dot. The
inset of Fig.~\ref{single} shows the occupation probabilities as a
function of time when $V^{o}_{w}$=5 meV. Note that for the whole
range of $V^{o}_{w}$ in Fig.~\ref{single} the final depth of the
SD is $V^{f}_{w}$=11 meV and $t_{a}=0.39T=0.14$ ns,
$t_{b}=0.58T=0.21$ ns. The loading mechanism is efficient since it
can induce a high loading probability i.e. $P_{SD}>$0.99 for
$V^{o}_{w}\lesssim2$ meV. By increasing $V^{o}_{w}$ higher
eigenstates mediate the tunneling process resulting in smaller
$P_{SD}$ since the width of the tunnel barrier between SD and
SAW-dot increases and inhibits electron tunneling into SD-states
which can follow LZTs. For the higher excited SD states the
process is not efficient because the electron tunnels in, and
subsequently off, the SD without following LZTs. For a deeper
$V^{f}_{w}$ additional excited states can be made to follow LZTs,
and thus to enhance the range for which $P_{SD}$ is close to
unity. We have also verified the robustness of the loading
mechanism for cases in which the initial SAW-dot state is a random
superposition of the low-lying SAW-eigenstates~\cite{note2}. We
have further verified that a second SAW-electron (which is in the
ground state of the SAW-dot) may be loaded successfully in the SD
when the first electron is assumed to be in the SD ground state
and $V_w(t)$ is increased according to Eq.(2). In principle, this
can happen for both singlet and triplet states, however,
$V^{f}_{w}$ must be deeper than for the loading of the first
electron since the second electron has to overcome the mutual
Coulomb repulsion with the first. In this way, the SD may be
loaded with two electrons which, after relaxation to the singlet
ground state, becomes the starting point for EPR pair generation
described earlier.

In Fig.~\ref{saw} we extend the results to two electrons carried
at $t_{o}=0$ in a single SAW-dot in the corresponding singlet
ground state, whose orbital distribution
$\rho=\int|\Psi(x,x')|^{2}dx'$ is shown in the inset of
Fig~\ref{saw}(a). This may be used, for example, as an alternative
to sequential loading of the SD prior to EPR pair generation.
Similarly to one electron $V_w(t)$ is increased according to
Eq.(2) and as we see in Fig~\ref{saw}(a) an increase of
$V^{o}_{w}$ lowers the occupation probability $P_{SD,SD}$, while
$P_{SAW,SAW}\sim0$ indicating that for this range of $V^o_{w}$
there is always enough time for one electron to tunnel into the SD
and follow LZTs. For the case $V^o_{w}$=2 meV Fig.~\ref{saw}(b)
shows the final electron distribution and Fig.~\ref{saw}(c) shows
the occupation probabilities as a function of time. Note that for
the whole range of $V^{o}_{w}$ in Fig.~\ref{saw} the final depth
of the SD is $V^{f}_{w}$=14.5 meV and $t_{a}=0.42T=0.16$ ns,
$t_{b}=0.7T=0.26$ ns. In general, the Coulomb repulsion between
the electrons in the SAW-dot enhances the tunneling process into
the SD for the first  electron, but the second electron has then
to overcome the Coulomb repulsion of the first, as with sequential
loading described above. If $V^{o}_{w}$ is sufficiently deep, the
second electron tunnels in and off the SD [right-hand side of
Fig.~\ref{saw}(a)] and, as excited states mediate the process, the
electron remaining in the SAW-dot is left in a superposition of
the lowest SAW-eigenstates. Unlike the narrow SD, the potential
minimum of the SAW-dot is relatively wide and as a result the
Coulomb energy dominates over the kinetic energy and hence the two
electrons are strongly correlated. For this reason they occupy
relatively distinct regions in the SAW-dot, as shown in the inset
of Fig.~\ref{saw}(a) and in fact both singlet and triplet states
have almost identical distributions yielding a small
antiferromagnetic exchange energy. This small energy splitting of
singlet and triplet leads to a high uncertainty in the degree of
entanglement of the incident electrons in the SAW-dot which in
turn leads to a high uncertainty in the resulting static-flying
qubit pair. This would not, therefore, be such an effective procedure
for producing fully entangled EPR pairs, as the method described
earlier which ensures relaxation to the singlet ground state
within the SD due to the much higher energy splitting of singlet
and triplet. However, we have proved that by tuning the depth of
the SD synchronously with the moving SAW-dot containing two
electrons, these electrons may both be bound in the SD with high
probability, as shown in Fig.~\ref{saw}(a), confirming this as an
alternative two-electron loading strategy of a SD.

Finally, we point out that the electron spins are subjected to
decoherence as in all solid state systems. The spin lifetime in
GaAs is $\sim$100 ns arising primarily from phonon
scattering\cite{kikkawa}. In our proposal the time scales of
interest of $\sim$ 0.4 ns as seen from Figs.~\ref{dot}(c) and
~\ref{saw}(c), are more than two orders of magnitude shorter.
Decoherence and error mechanisms for relevant SAW-based devices
have been described in Refs.~16 and ~17 and include for example
coupling of electron spins to nuclear spins\cite{petta}, noise on
surface gates, impurities and disorder, errors in the width of the
depleted channel and temperature effects. A detailed study of
decoherence effects for our scheme will be presented elsewhere.

In conclusion, we have proposed and studied an electron spin
entangler which generates in a controlled way nonlocal singlet
states and can be realised with existing semiconductor technology.
The entangler scheme utilizes a synchronized time dependent gate
voltage which tunes the confining potential of a SD formed in a
depleted channel, and a SAW-dot which propagates through the
channel and captures (transfers) electrons from (to) the SD. By
studying the exact time evolution of two-electron states we showed
that static-flying EPR electron spin pairs, which may be converted
to static or flying, can be efficiently produced within times of
0.4 ns. We also demonstrated a SAW scheme of loading a SD with one
and two electrons, enabling arbitrary interconversion of flying
and static spin qubits bound in a quatum dot.

We thank M. Kataoka for discussions. This work was supported by
the UK MoD and is part of the UK QIP IRC (www.qipirc.org,
GR/S82176/01).


\begin{thebibliography}{35}
\bibitem{note0}
The occupation probabilities shown in this work have meaning only
when the SAW-dot and SD regions do not overlap. This is exactly
the case at the final time $t_{f}=T=0.37$ ns, that we are interested
in, when the probabilities have obtained their asymptotic values.
\bibitem{note1}
In principle, the number of electrons in the SAW-dots can be
controlled accurately with the gate voltage which is applied to
form the channel along which the SAW propagates (see for example
Refs.~13 and 14).
\bibitem{note2}
This is useful if we want to convert the static-flying EPR pair to
a static pair, because as we saw in Fig.~\ref{dot}(b) the electron
which tunnels from the SD to the SAW-dot is in a superposition of
SAW-eigenstates.
\end{thebibliography}
\end{document}